\documentclass{article}

\PassOptionsToPackage{numbers, compress, sort}{natbib}

\usepackage[dandb,final]{neurips_2025}

\newcommand{\eat}[1]{}

\usepackage{balance}
\usepackage{times}
\usepackage{latexsym}
\usepackage{amsfonts}
\usepackage{amsmath}
\usepackage{colortbl}
\usepackage{epsfig}
\usepackage{xspace}
\usepackage{graphicx}
\usepackage{comment}
\usepackage{booktabs}
\usepackage{multirow}
\usepackage{framed}
\usepackage{subcaption}

\usepackage{circledsteps}
\usepackage[utf8]{inputenc}
\usepackage{microtype}
\usepackage{inconsolata}
\usepackage[colorlinks=true, linkcolor=black, urlcolor=magenta, citecolor=black]{hyperref}

\usepackage[ruled,linesnumbered]{algorithm2e}
\usepackage{algpseudocode}

\algnewcommand\algorithmicinput{\textbf{Input:}}
\algnewcommand\algorithmicoutput{\textbf{Output:}}
\algnewcommand\Input{\item[\algorithmicinput]}
\algnewcommand\Output{\item[\algorithmicoutput]}

\algdef{SE}[DOWHILE]{Do}{doWhile}{\algorithmicdo}[1]{\algorithmicwhile\ #1}%

\definecolor{green}{RGB}{0,128,0}

\definecolor{yellow}{RGB}{255,200,18}

\newcommand{\stab}{\vspace{1.2ex}\noindent}

\newcommand{\bi}{\begin{itemize}}
\newcommand{\ei}{\end{itemize}}

\newcommand{\be}{\begin{enumerate}}
\newcommand{\ee}{\end{enumerate}}
\newcommand{\beqn}{\begin{eqnarray*}}
\newcommand{\eeqn}{\end{eqnarray*}}

\newcommand{\stitle}[1]{\stab\noindent{\bf #1}}
\newcommand{\etitle}[1]{\vspace{1mm}\noindent{\underline{\em #1}}}
\newcommand{\ie}{{\em i.e.,}\xspace}
\newcommand{\eg}{{\em e.g.,}\xspace}

\usepackage{tabu}                      
\usepackage{booktabs}                  
\usepackage{lipsum}                    
\usepackage{mwe}                       

\usepackage{listings}
\usepackage{tcolorbox}
\usepackage{pifont}
\usepackage{tikz}
\usepackage{enumitem}
\usepackage{natbib}
\usepackage{utfsym} 

\usetikzlibrary{shapes,snakes}
\usetikzlibrary{calc}

\definecolor{shadecolor}{RGB}{220,220,220}

\tikzstyle{mybox} = [draw=black, fill=black!5, thick,
    rectangle, rounded corners, inner sep=2pt, inner ysep=8pt]
\tikzstyle{fancytitle} =[fill=black, text=white]

\newcommand{\sys}{DeepFund\xspace}

\title{Time Travel is Cheating: Going Live with DeepFund for Real-Time Fund Investment Benchmarking}

\author{%
Changlun Li$^{1,2,}$\thanks{Both authors contributed equally to this paper} ,
~Yao Shi$^{1,2,*}$, 
~Chen Wang, 
~Qiqi Duan, 
~Runke Ruan, \\
\bf
Weijie Huang,
~Haonan Long, 
~Lijun Huang, 
~Nan Tang$^{1,2}$,
~Yuyu Luo$^{1,2,}$\thanks{Yuyu Luo is the corresponding author (yuyuluo@hkust-gz.edu.cn)}
\\
$^1$The Hong Kong University of Science and Technology (Guangzhou) \\
$^2$Paradoox AI Research \\
}

\begin{document}

\maketitle

\begin{abstract}
Large Language Models (LLMs) have demonstrated notable capabilities across financial tasks, including financial report summarization, earnings call transcript analysis, and asset classification. However, their real-world effectiveness in managing complex fund investment remains inadequately assessed. A fundamental limitation of existing benchmarks for evaluating LLM-driven trading strategies is their reliance on historical back-testing, inadvertently enabling LLMs to ``time travel'' -- leveraging future information embedded in their training corpora, thus resulting in possible information leakage and overly optimistic performance estimates. To address this issue, we introduce \sys, a live fund benchmark tool designed to rigorously evaluate LLM in real-time market conditions. Utilizing a multi-agent architecture, \sys connects directly with real-time stock market data -- specifically data published after each model’s pretraining cutoff -- to ensure fair and leakage-free evaluations. Empirical tests on nine flagship LLMs from leading global institutions across multiple investment dimensions—including ticker-level analysis, investment decision-making, portfolio management, and risk control—reveal significant practical challenges. Notably, even cutting-edge models such as DeepSeek-V3 and Claude-3.7-Sonnet incur net trading losses within \sys real-time evaluation environment, underscoring the present limitations of LLMs for active fund management. Our code is available at \url{https://github.com/HKUSTDial/DeepFund}.
\end{abstract}

\section{Introduction}

The financial industry has witnessed an AI-driven revolution over the past decade~\cite{li2023large,Liu2024ASO,zhao2024revolutionizing,wang2025quantbench}. Advanced AI techniques, particularly Large Language Models (LLMs), have transformed practices across multiple domains, including high-frequency trading algorithms~\cite{Liu2020AdaptiveQT,Briola2021DeepRL,Zong2024MacroHFTMA,liu2020finrl}, risk assessment models~\cite{giudici2018fintech,zheng2019finbrain,javaid2024ai}, investment decisions~\cite{xiao2024tradingagents,li2024cryptotrade,yu2024finmem,zhang2024multimodal,li2025hedgeagents}, and data analysis~\cite{verifyai,nl2sql360, DBLP:journals/tvcg/LuoTLTCQ22, DBLP:journals/tkde/LiuSLMJZFLTL25,DBLP:journals/pvldb/LuoLFCT25, DBLP:journals/pvldb/LiYLFT25}, thereby fundamentally reshaping how financial institutions operate and make decisions.

\begin{figure*}[t!]
    \vspace{-1em}	
	\centering
	\includegraphics[width=\textwidth]{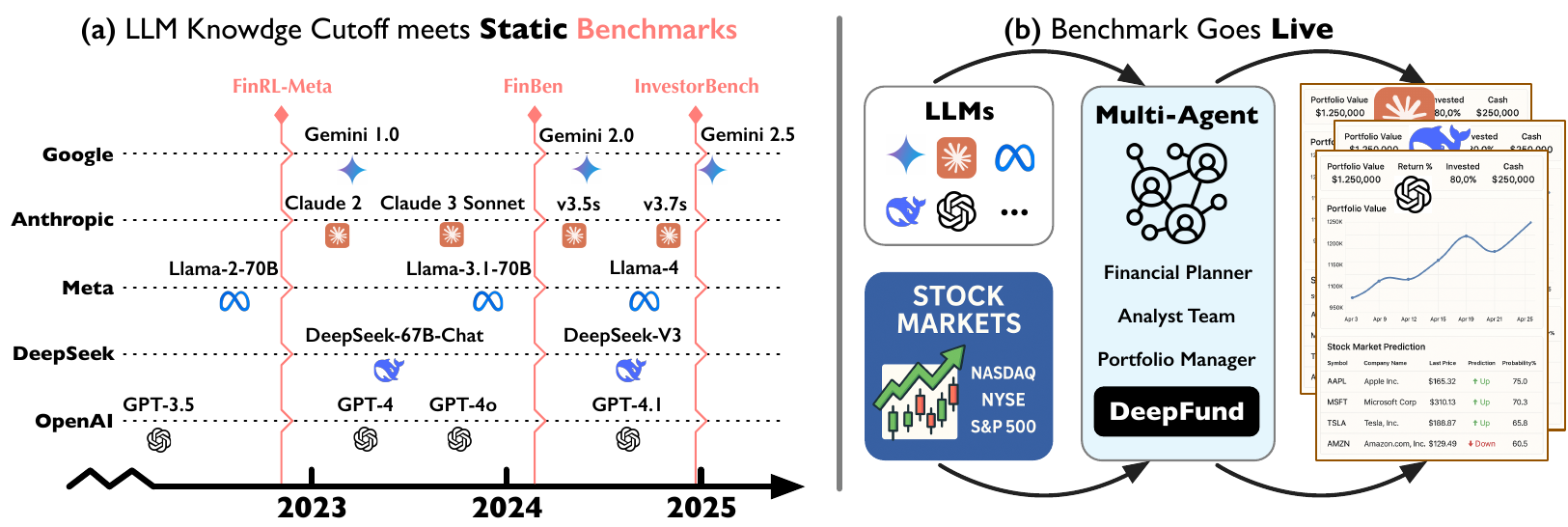}
 	\caption{Shifts from static benchmarks to live benchmarks. Particularly for relevant date information in 1(a), we refer to public sources (\eg model card, arXiv, GitHub) for illustration.}
	\label{fig:dilemma}
    \vspace{-1em}	
\end{figure*}

Current financial benchmarks—such as TAT-QA~\cite{Zhu2021TATQAAQ}, FinanceBench~\cite{islam2023financebench}, FinBen~\cite{xie2024finben}, and InvestorBench~\cite{li2024investorbench}—have made valuable contributions by assessing an LLM’s understanding of financial documents, terminology, and trading performance. 
For example, these benchmarks typically evaluate LLMs' effectiveness by simulating trading strategies using historical market data, measuring performance based on metrics such as cumulative returns or risk-adjusted returns.
Such evaluations have been widely adopted for both general-purpose foundation models and finance-specific LLMs. \textbf{However, a critical gap is that these benchmarks primarily probe static data rather than a model’s ability to make effective investment decisions in real-time market conditions}.

A fundamental limitation in extending these evaluations to \textbf{trading performance} lies in their reliance on retrospective back-testing. Back-testing is the standard method for assessing trading strategies~\cite{Arnott2018ABP,Harvey2015Backtesting}, but it becomes problematic when applied to LLM-driven strategies because the model may have been pre-trained on the very historical data used for testing, which leads to a severe \textbf{information leakage issue}~\cite{Xu2024BenchmarkingBL,ravaut2024much,ni2025training,dekoninck2024constat}. Undoubtedly, an LLM can appear to perform extraordinarily well on historical market data simply by regurgitating events it has already seen, rather than genuinely predicting outcomes~\cite{lopez2025memorization,Glasserman2023AssessingLB,llmstat,zhu2025elliesql}. In other words, the model can effectively ``\textbf{time travel}'' by using future knowledge during evaluation—a form of cheating that inflates its apparent performance. This issue is exacerbated by the varying knowledge cut-off dates of different LLMs. 

As shown in Figure~\ref{fig:dilemma}(a), GPT-4o was trained on data up to October 2023~\cite{hurst2024gpt}, whereas DeepSeek-V3’s training extends until July 2024~\cite{liu2024deepseek}. If we evaluate such a model on a period prior to its knowledge cutoff (\eg testing DeepSeek-V3 on 2021–2023 data), it will have effectively already seen those market conditions during pre-training, yielding overly optimistic metrics that do not reflect true predictive power. 

\stitle{\sys: Fund Benchmark Going Live.} To address the critical gap identified above, we introduce \sys – a comprehensive framework for real-time fund investment benchmarking, as shown in Figure~\ref{fig:dilemma}(b). Inspired by previous works~\cite{liu2022finrl_meta,jain2024livecodebench,colin2025livebench,zhuang2025large,DBLP:conf/sigmod/Luo00CLQ21}, \sys assesses LLM's ability to make effective investment decisions in a live-market environment, explicitly preventing any leakage of future data. In particular, our approach offers three key contributions:

(a) \textbf{\textit{Live Forward Testing:}} We introduce a novel benchmarking tool that supports real-time trading conditions to mitigate information leakage. Meanwhile, we provide an interactive web-based interface for performance visualization and comparative analysis on domain-specific financial metrics (\eg Cumulative Return, Sharpe Ratio) to rigorously assess LLMs' effectiveness as fund managers.

(b) \textbf{\textit{Multi-Agent Decision Framework:}} We implement a multi-agent architecture in which LLMs assume multiple roles (acting as financial planner, analyst team, and portfolio manager), thereby creating a realistic reproduction of the investment decision-making process. This design mirrors how human analysts and portfolio managers collaborate.

(c) \textbf{\textit{Empirical Findings:}} Through rigorous live environment interaction with various LLMs, we reveal significant performance disparities, highlighting the challenges and possibilities of LLMs in real-time trading, and demystify distinct trading behaviors and personalities exhibited by different LLMs.





\section{DeepFund: Multi-Agent Fund Investment Going Live}

\begin{figure}
	\centering
	\includegraphics[width=\textwidth]{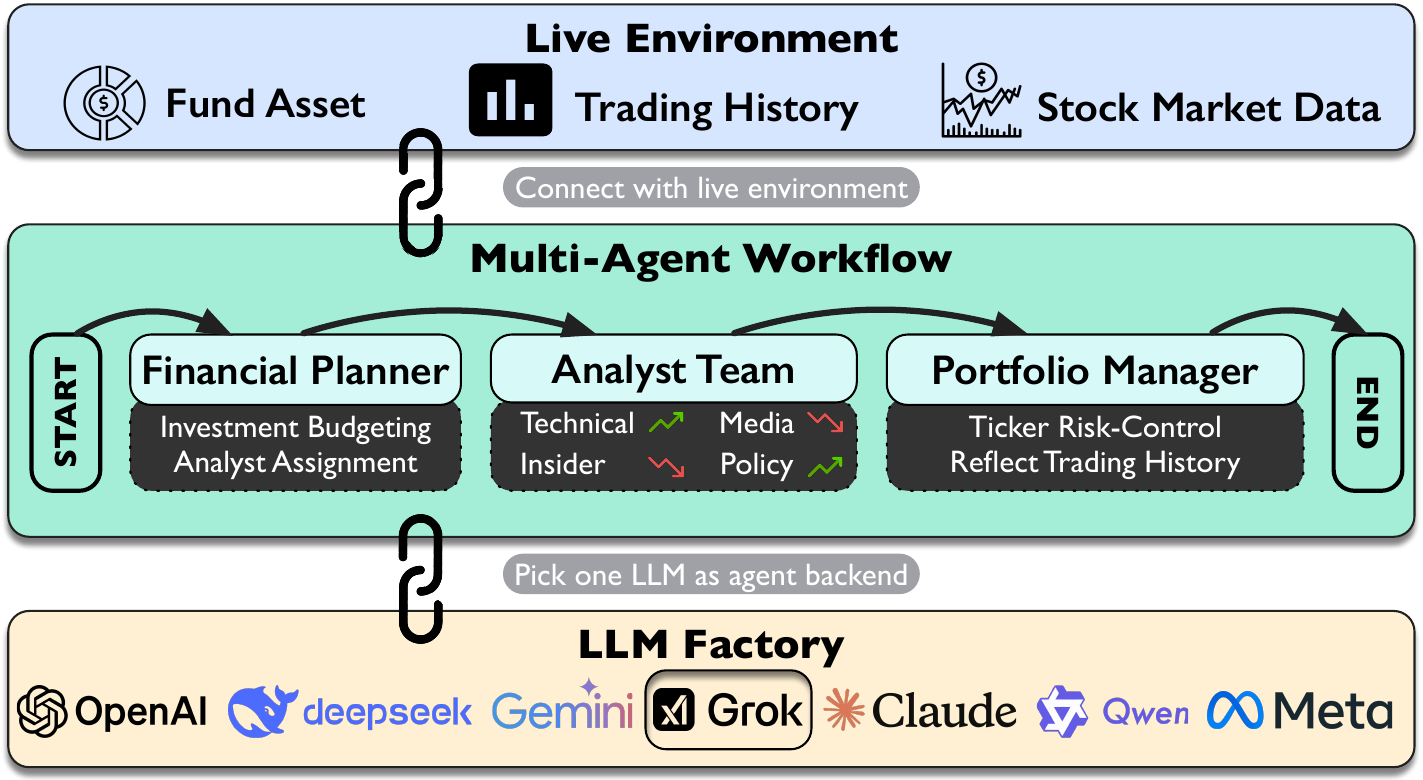}
	\caption{The \sys framework.}
	\label{fig:deepfund}
    \vspace{-2em}	
\end{figure}

\sys is designed to emulate the dynamics of a real-world fund investment environment, as illustrated in Figure~\ref{fig:deepfund}.
At the top, the \textbf{Live Environment} continuously ingests real-time market data, fund asset information, and trading history, ensuring realistic conditions free from information leakage.
Below, the \textbf{Multi-Agent Workflow} mimics a structured fund management process through three distinct roles: \textit{\textbf{Financial Planner}}, \textit{\textbf{Analyst Team}}, and \textit{\textbf{Portfolio Manager}}.
The entire workflow is powered by a single LLM, selected from various providers available in the \textbf{LLM Factory} (\eg Grok as selected in Figure~\ref{fig:deepfund}), ensuring flexible and consistent backend capabilities across all agents.

%

\stitle{Live Environment.}
The live environment serves as the cornerstone of \sys, enabling  real-time market conditions essential for robust and leakage-free evaluation. Unlike traditional static benchmarks, our environment continuously integrates dynamic financial data streams from three distinct sources: (1) real-time \textbf{stock market data}, providing immediate market movements and price fluctuations; (2) up-to-date \textbf{fund asset} information, reflecting the current state of investment positions; and (3) detailed \textbf{trading history}, recording all activities related to portfolio management.

To facilitate seamless and flexible data ingestion, we implement a \textbf{modular API gateway} that interfaces with multiple financial data providers, such as Yahoo Finance and Alpha Vantage. This modularity (refer to Appendix~\ref{appendix:scalability}) ensures adaptability to diverse data sources and straightforward integration into varying market contexts. By offering direct feedback from live market conditions, the environment guarantees authenticity in the decision-making process, fundamentally shifting the evaluation paradigm from retrospective back-testing to dynamic, real-time interaction.

\stitle{Single Agent Design.}
Similar to previous works~\cite{li2024cryptotrade,xiao2024tradingagents,li2025hedgeagents,luo2025nvbench20resolvingambiguity,10.1145/3711896.3737427}, each agent in our framework, powered by the selected LLM backend, fulfills a specific role within the investment process:

\etitle{Financial Planner:} Strategically orchestrates the investment analysis by determining analytical priorities and allocating tasks to suitable analysts. It supports two modes: a deterministic mode, allowing predefined analyst selection, and a dynamic mode, leveraging self-reasoning to flexibly select analysts.

\etitle{Analyst Team:} Consists of specialized analyst agents—Fundamental, Technical, Insider, Company News, Macro Economic, and Policy—that analyze domain-specific data and generate standardized signals \texttt{(Bullish, Bearish, or Neutral)}, accompanied by detailed justifications.
See Table~\ref{tab:analysts} for analyst types and their specialized functions.

\etitle{Portfolio Manager:} Integrates multiple analyst signals to make executive investment decisions \texttt{(Buy, Sell, Hold)}, manages risk control (\ie the portion of holdings and cash), and maintains a dual-memory architecture (see Appendix~\ref{appendix:memory}) to reflect historical transactions and current portfolio states.

\begin{table}[t!]
	\caption{Analyst types, their data sources, and specialized features within the Analyst Team.}
	\label{tab:analysts}
	\centering
	\begin{tabular}{p{0.2\linewidth}p{0.2\linewidth}p{0.5\linewidth}}
	\toprule
	\textbf{Analyst Type} & \textbf{Data Source} & \textbf{Feature} \\
	\midrule
	Technical & Historical price/volume data & Focuses on price patterns and indicators, such as trends, RSI, volatility, support/resistance. \\
	Fundamental & Financial statements, ratios & Analyzes company financials, such as earnings, margins, valuation metrics. \\
	Insider & Insider transaction reports & Monitors corporate insider activity, such as executive buys/sells, timing patterns. \\
	Company News & News articles, press releases & Assesses company-specific news, such as sentiment, material events. \\
	Macro Economic & Economic indicators & Examines economic conditions, such as GDP, inflation, unemployment, rates. \\
	Policy & Policy news, central bank reports & Analyzes fiscal/monetary policy, such as interest rates, spending, regulation. \\
	\bottomrule
	\end{tabular}
\end{table}

\stitle{Multi-Agent Workflow.}
The multi-agent workflow~\cite{aflow,liu2025advances,wang2024agent,DBLP:conf/icde/LuoQ0018,alphasql} adopts an orchestrator-worker paradigm\footnote{See Anthropic blog \url{https://www.anthropic.com/engineering/building-effective-agents}} to mimic a realistic fund management process. Initially, the \textbf{Financial Planner} selects and assigns analysts based on real-time market conditions and portfolio status. Next, the selected \textbf{Analyst Team} concurrently processes domain-specific information and generates structured analytical signals. Finally, the \textbf{Portfolio Manager} synthesizes these signals, evaluates portfolio risks, decides optimal trading actions, and updates the investment portfolio state. The entire process is rigorously tracked, ensuring complete traceability and consistency throughout the decision-making pipeline. We provide more details on how the workflow effectively coordinates information exchange in Appendix~\ref{appendix:schema}.

\stitle{Evaluation Interface.} As shown in Figure~\ref{fig:dilemma}(b), we provide a web-based interface for presenting the trading performance for each LLM. Inspired by the previous work, ChatBot Arena~\cite{chiang2024chatbot} and Open FinLLM Leaderboard~\cite{lin2025open}, the interface is designed to be comprehensive and fine-grained, allowing for in-depth analysis of the trading behavior. Please refer to Appendix~\ref{appendix:interface} for more details.

\section{Experimental Setting}
\label{sec:setting}

\stitle{Financial Data Integration.}
We integrate upstream data from well-known and trusted financial provider APIs (\eg Alpha Vantage, Yahoo Finance). The data covers not only granular ticker-level information, such as financial statement, company news, daily trading statistics, and insider transactions, but also the macro indicators and policy news. Parametric settings are presented in Appendix~\ref{appendix:parameter} for more details.

\stitle{LLMs.} 
We evaluated nine state-of-the-art LLMs from various providers, each with distinct knowledge cutoff dates, to fairly assess their real-time investment performance, as shown in Table~\ref{tab:llms}.


\begin{table}[t!]
	\caption{Evaluated LLMs: The detailed information is sourced from related technical report.}
	\label{tab:llms}
	\centering
	\begin{tabular}{@{}ccccc@{}}
	\toprule
	\textbf{Provider} & \textbf{Model Version} & \textbf{Open Source} & \textbf{Release Date} & \textbf{Knowledge Cutoff} \\ \midrule
	OpenAI            & GPT-4.1~\cite{openai2025gpt41}                & ✗   & Apr 2025  & June 2024 \\
	Meta              & Llama 4 Scout~\cite{meta2025llama4scout}          & ✓   & Apr 2025  & Aug 2024  \\
	Google            & Gemini 2.5 Flash~\cite{google2025gemini25flash}       & ✗   & Apr 2025  & Jan 2025  \\
	Anthropic         & Claude 3.7 Sonnet~\cite{anthropic2025claude37sonnet}      & ✗   & Feb 2025  & Oct 2024  \\
	xAI               & Grok 3 mini Beta~\cite{xai2025grok3minibeta}       & ✗   & Feb 2025  & Nov 2024  \\
	DeepSeek          & DeepSeek-V3~\cite{deepseek2025deepseekv3}       & ✓   & Mar 2025  & Dec 2024  \\
	Alibaba           & Qwen2.5-Max~\cite{alibaba2025qwen25max}           & ✗   & Jan 2025  & NA         \\
	ByteDance         & Doubao-1.5-pro~\cite{byte2025doubao15pro}         & ✗   & Jan 2025  & NA         \\
	Zhipu             & GLM-4-Air~\cite{zhipu2025glm4air}              & ✓   & Apr 2025  & NA         \\ \bottomrule
	\end{tabular}
\end{table}

\stitle{Portfolio Configuration.}
In a nod to the investment wisdom of Warren Buffett, our experiments target investments in Berkshire Hathaway’s top five holdings as of Q1 2025: Apple (AAPL), American Express (AXP), Bank of America (BAC), Coca-Cola (KO), and Chevron (CVX). Each LLM manages initial cashflow with a total amount of \$100,000.
As the \emph{Fundamental} and \emph{Macro-Economic} analysts are designed to provide long-term perspective (\ie quarterly, half-yearly, or annually), to accommodate the daily trading frequency, each LLM will coordinate the other four analysts: \emph{Technical}, \emph{Company News}, \emph{Policy}, and \emph{Insider} in our experiments.

\stitle{Trading Period.}
The trading period is from March 17 to April 17, 2025, covering 24 trading days with a daily trading frequency. Notably, this period captures two significant market events:
(1) \textbf{FOMC Meeting}: During March 18-19, the Federal Reserve maintained the federal funds rate at 4.5\%, marking a second consecutive pause following earlier rate cuts. 
(2) \textbf{Tariff Impact}: During April 2-9, the US government first announced a heavy tariff on global imports, then paused it. This move intensified concerns over inflation and economic slowdown, contributing to fluctuations in major stock indices.

\stitle{Signal and Decision Validity.}
The validity of signals and decisions generated by \sys was evaluated over 24 trading days. The signal and decision are regarded as valid only if the justification is correctly provided. Failing to do so will result in \emph{No signal provided due to error} or \emph{Just hold due to error}, respectively. Out of a total of 4320 signals and 1080 trading decisions, \sys successfully produced \textbf{4144 signals (96\% validity) and 1059 trading decisions (98\% validity)}. Such high validity rates indicate the robustness and reliability in generating timely and actionable outputs. Further detailed statistics are available in Appendix~\ref{appendix:key_stats}.

\stitle{Evaluation Metrics.} We employ standard financial metrics to measure performance rigorously, including Cumulative Return ($\mathbf{CR}$)~\cite{hull2012risk}, Cumulative Return at Buy \& Hold ($\mathbf{CR}_{bnh}$)~\cite{fama1988permanent}, Sharpe Ratio ($\mathbf{SR}$)~\cite{sharpe1994sharpe}, Maximum Drawdown ($\mathbf{MDD}$)~\cite{magdon2004maximum}, Win Rate ($\mathbf{WR}$)~\cite{chan2009quantitative}, Beta ($\beta$)~\cite{jensen1968performance}, and Alpha ($\alpha$)~\cite{jensen1968performance}. The detailed definitions and formulas for these metrics can be found in Appendix~\ref{appendix:metric}.

\stitle{Implementation Details.} 
We build the agentic workflow via LangChain, a powerful toolkit for building LLM-based applications.
All LLM inferences occur through provider-specific APIs, using standardized prompts (see Appendix~\ref{appendix:prompt}) and the same temperature for fairness (refer to Appendix~\ref{appendix:parameter} for more details).
Additionally, we utilize Supabase, a PostgreSQL-based cloud database, to store all activities, including historical decisions, portfolio states, and analytical signals.
The evaluation incurred approximately \$100 in total costs (\ie LLM APIs 40\%, financial data 40\%, and cloud database 20\%, correspondingly). To support reproducibility, our code repository is publicly available.



\section{Going Live: Revealing the True Trading Power of LLMs}


In this section, we delve into the real-world trading performance and behavior of various LLMs deployed in a live market environment. We first present a comprehensive overview of the trading outcomes for all evaluated LLMs. Subsequently, we conduct an in-depth comparative analysis of two representative models: Grok, which uniquely achieved profitability, and DeepSeek, which experienced losses. Through detailed observation of their underlying reasoning processes and decision chains within our trading pipeline, we aim to uncover the critical factors contributing to their divergent performance. Moreover, we provide additional intriguing findings in Appendix~\ref{appendix:add_exp}.

This analysis seeks to answer pivotal questions regarding the practical capabilities of LLMs in financial markets:
\textbf{Q1:} Which LLMs thrive—and which struggle—in the high-stakes arena of live trading?
\textbf{Q2:} How adeptly can LLMs transform complex, multi-source financial data into precise and actionable trading signals?
\textbf{Q3:} When the signals are translated into real market actions, do LLMs truly achieve profitable outcomes?
\textbf{Q4:} What unique trading ``personalities'' or strategic styles can we uncover among different LLMs in the live market setting?

\subsection{Beyond Backtesting: LLMs Face the Live Market Challenge (Q1)}

\begin{figure}
    \centering
    \includegraphics[width=\textwidth]{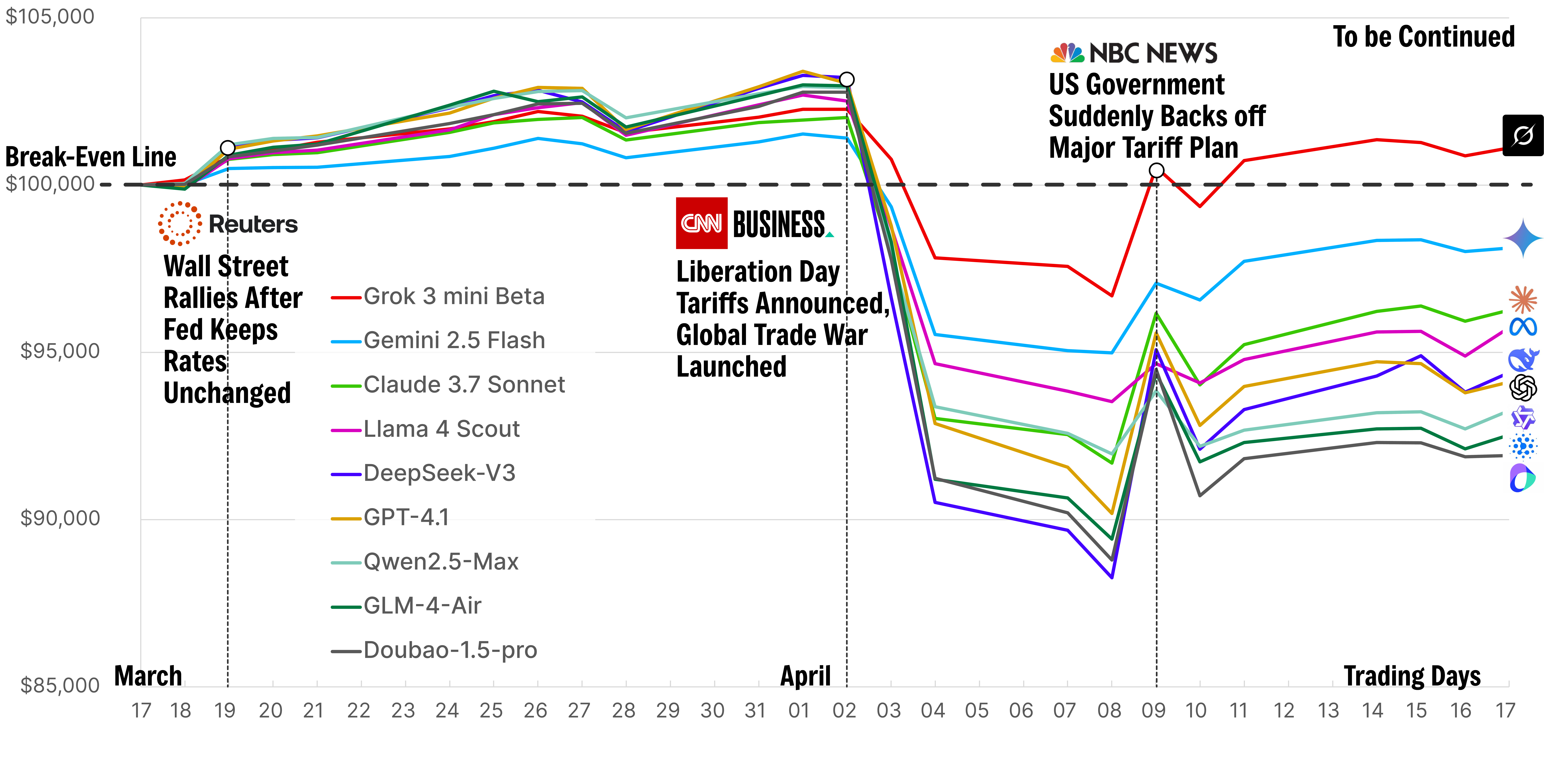}
    \vspace{-3em}
    \caption{Portfolio asset value for each LLM over time.}
    \label{fig:total_assets}
    \vspace{-1em} 
\end{figure}

The financial markets are notoriously challenging, often echoing the old adage, ``Out of ten gamblers, nine will lose''. Our live trading experiment with LLMs starkly underscores this reality. Table~\ref{tab:performance} and Figure~\ref{fig:total_assets} reveal that while all evaluated LLMs were capable of executing the end-to-end ``data-signal-decision'' pipeline, their profitability varied dramatically under identical market conditions. A significant majority experienced net trading losses (\ie Cumulative Return $\mathbf{CR}<0$), highlighting the substantial hurdles of achieving success in real-time fund investment. Strikingly, only the Grok 3 model managed to secure a positive cumulative return. 

After the FOMC meeting, all LLMs kept positive net gain. When it came to the tariffs impact, we observe that the DeepSeek incurred the largest drawdown, which is 14.5\%. During the bearish market period, Table~\ref{tab:performance} shows that most US-produced LLMs (except for GPT-4.1) demonstrated lower return losses (\ie $\mathbf{CR}$) than Chinese-produced LLMs. Compared to the $\mathbf{CR}_{bnh}$, we observe that a passive Buy \& Hold strategy would have more resilience to such market fluctuations.

\begin{tikzpicture}
\node [mybox] (box){%
    \begin{minipage}{\columnwidth}
    {
    The Live Market Gauntlet: Most LLMs Stumble, but Grok Emerges as the Lone Survivor!
    }
    \end{minipage}
};
\node[fancytitle, rounded corners] at (box.south) {{\bf Q1 Takeaway}};
\end{tikzpicture}



\begin{table}[t!]
    \caption{Overall trading performance of LLMs in DeepFund, sorted by $\mathbf{CR}$ ($\downarrow$).}
    \label{tab:performance}
    \centering
    \begin{tabular}{cccccccc}
    \toprule
    \textbf{Model Version} & \textbf{CR(\%)} & \textbf{CR$_{bnh}$ (\%)} & \textbf
    {SR} & \textbf{MDD} (\%) & \textbf{WR (\%)} & $\beta$ & $\alpha$ \\
    \midrule
    Grok 3 mini Beta & \cellcolor{green!10}\textbf{+1.1} & -3.09 & 0.51 & 5.5 & 61 & 0.42 & 0.2 \\
    Gemini 2.5 Flash & \cellcolor{red!10}-1.9 & -1.58 & -1.37 & 6.4 & 61 & 0.35 & 0.0 \\
    Claude 3.7 Sonnet & \cellcolor{red!20}-3.7 & -2.94 & -1.45 & 10.1 & 70 & 0.64 & 0.0 \\
    Llama 4 Scout & \cellcolor{red!30}-4.3 & -3.62 & -2.42 & 8.9 & 61 & 0.36 & -0.1 \\
    DeepSeek-V3 & \cellcolor{red!40}-5.7 & -5.6 & -1.39 & 14.5 & 57 & 0.94 & 0.0 \\
    GPT-4.1 & \cellcolor{red!50}-5.9 & -4.41 & -1.87 & 12.8 & 52 & 0.77 & 0.0  \\
    Qwen2.5-Max & \cellcolor{red!60}-6.7 & -4.86 & -3.12 & 10.7 & 65 & 0.48 & -0.2 \\
    GLM-4-Air & \cellcolor{red!70}-7.5 & -3.90 & -2.31 & 13.2 & 57 & 0.78 & -0.1 \\
    Doubao-1.5-pro & \cellcolor{red!80}-8.1 & -5.37 & -2.35 & 13.6 & 65 & 0.84 & -0.1 \\
    \midrule
    S\&P 500 & \cellcolor{yellow!80}-6.91 & NA & 0.3 & 13.7 & NA & 1.00 & 0.0 \\
    \bottomrule
    \end{tabular}
\end{table}

\subsection{Signal or Noise? Decoding LLMs' Analytical Powers (Q2)}

The system employs four functional analysts that generate signals \texttt{(Bullish}, \texttt{Neutral}, or \texttt{Bearish)} based on diverse inputs. The efficacy of an LLM-driven analyst in signal extraction is reflected in the alignment of its aggregated signals with subsequent stock price movements. To assess this, we bind live environment with multi-source data, including company news, insider transactions, policy shifts, and technical indicators. Each analyst is required to generate a signal and justify its reasoning. 

\begin{figure}[t!]
    \vspace{-2em}
    \centering
    \begin{subfigure}[b]{\textwidth}
        \centering
        \includegraphics[width=\linewidth]{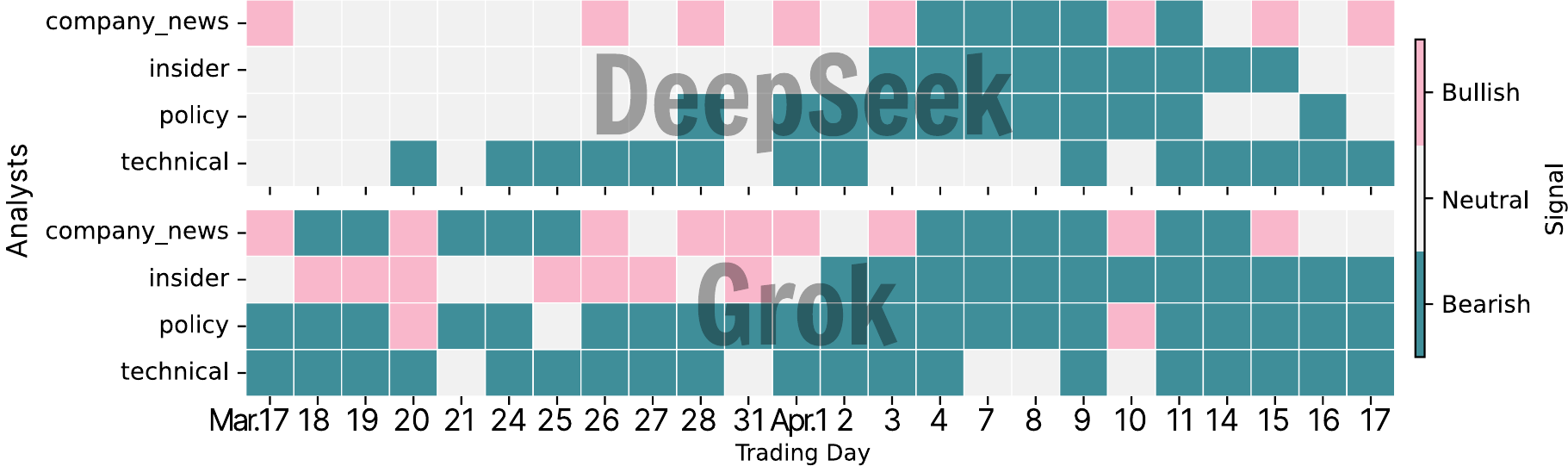}
        \caption{Daily analyst signal overview by DeepSeek(up) and Grok(down).}
    \end{subfigure}
    \hfill 
    \begin{subfigure}[b]{\textwidth}
        \centering
        \includegraphics[width=\linewidth]{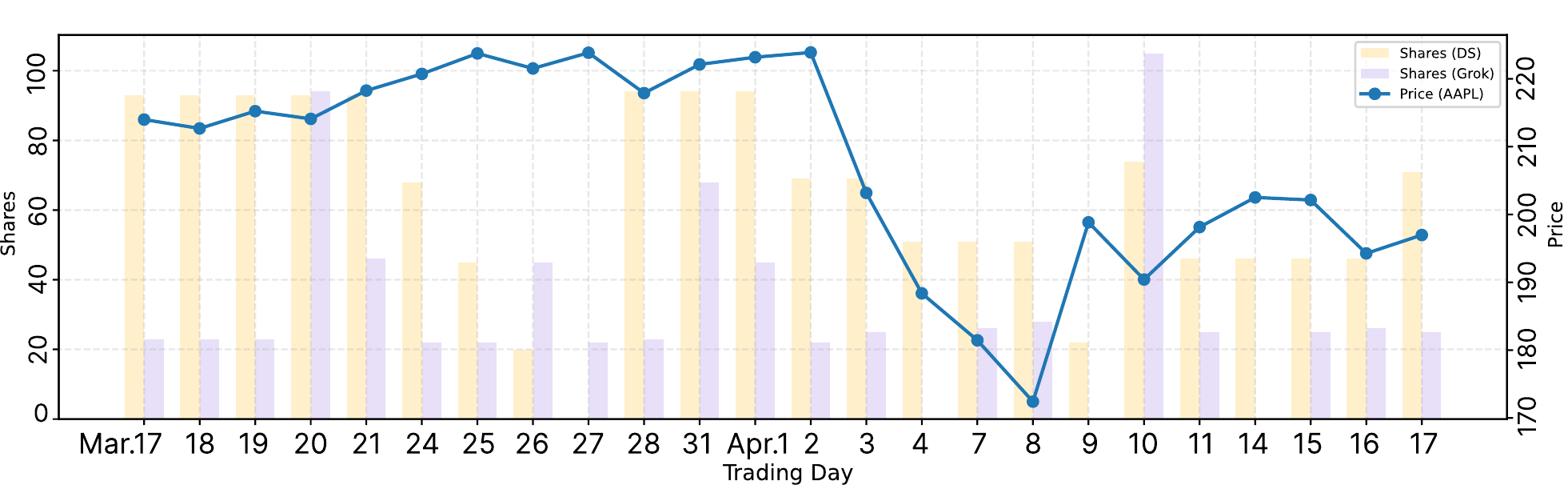}
        \caption{Stock price and holdings for DeepSeek (yellow) and Grok (purple) over the trading period.}
    \end{subfigure}
    \caption{AAPL trading for DeepSeek and Grok.} 
    \label{fig:aapl_grok_deepseek}
    \vspace{-2em}
\end{figure}

Figure~\ref{fig:aapl_grok_deepseek}(a) presents the analyst signal distributions for Apple Inc. (AAPL) powered by Grok and DeepSeek correspondingly. Generally, Grok produced a higher proportion of directional signals \texttt{(Bullish} or \texttt{Bearish)}, resulting in greater signal diversity. In contrast, DeepSeek leaned heavily towards \texttt{Neutral} signals given identical conditions.

Specifically, during the period of modest price fluctuation (March 17 to April 2), DeepSeek preferred \texttt{Neutral} signals, suggesting a less sensitive stance. While, Grok preferred \texttt{Bullish} signals. When the tariffs were announced, both turned to pump \texttt{Bearish} signals, indicating a more cautious stance.
Notably, both models struggled to predict the significant price surge on April 9 (from 172.42 USD to 198.85 USD), indicating a shared limitation in detecting strong reversal signals.

Viewing from the perspective of data sources, the behavior revealed further distinctions.
Grok-based analysts consistently displayed a bearish perspective in policy and technical analyses, except during tariff-influenced periods. An interesting divergence occurred on April 10 (refer to case study in Appendix~\ref{appendix:case_1} for details), where Grok and DeepSeek held opposing views in policy analysis: both acknowledged short-term uncertainty, but Grok expressed optimism regarding long-term prospects.

\begin{tikzpicture}
\node [mybox] (box){%
    \begin{minipage}{\columnwidth}
    {
    Reading the Market's Pulse: Grok Captures Policy Shifts and Technical Trends, DeepSeek Stays Neutral and Misses Key Signals!
    }
    \end{minipage}
};
\node[fancytitle, rounded corners] at (box.south) {{\bf Q2 Takeaway}};
\end{tikzpicture}



\subsection{From Signals to Profits: The Real Test of Trading Decisions (Q3)}

In this stage, LLM ingests collective signals, trading history, and holding shares to make trading decisions. 
We evaluated this signal-to-decision consistency by analyzing how well aggregated signals turned into actions. Intuitively, the ``Signal-to-Decision'' is regarded as consistent if observed:
(i) a \texttt{Buy} followed dominant \texttt{Bullish} signals;
(ii) a \texttt{Sell} followed dominant \texttt{Bearish} signals;
(iii) a \texttt{Hold} occurred with \texttt{Neutral} or mixed signals without a clear directional bias.

Figure~\ref{fig:aapl_grok_deepseek}(b) shows AAPL's price movements and the trading positions of Grok and DeepSeek. Both models generally integrate signals with the following evidence:
(i) \texttt{Sell} decisions are often aligned with \texttt{Bearish} signal from technical and policy analysts;
(ii) both models tend to \texttt{Hold} with mixed or predominantly \texttt{Bearish} signals;
(iii) insider source has limited impact on \texttt{Buy} and \texttt{Sell} choices. 

Particular to the \texttt{Buy} decisions, Grok was more influenced by \texttt{Bullish} signals from company news and policy, while DeepSeek weighted company news more heavily. Although both showed decision-making consistency, Grok demonstrated superior information integration by incorporating policy signals and exhibiting greater decision flexibility, suggesting better market timing capabilities.

Crucially, consistency does not guarantee effectiveness in profitability. We defined an effective decision as a \texttt{Buy} followed by a price increase or a \texttt{Sell} followed by a decrease (excluding the final day). Grok made 11 \texttt{Buy} decisions (7 effective) and 10 \texttt{Sell} decisions (5 effective). Effective decisions are often correlated with \texttt{Bearish} policy/technical or \texttt{Bullish} company news signals. DeepSeek made 3 \texttt{Buy} decisions (1 effective) and 8 \texttt{Sell} decisions (5 effective), showing higher precision in sells but fewer effective buys. Overall, Grok was more effective in leveraging diverse signals, while DeepSeek cautious stance, though precise in sells, limited its ability to capitalize on market opportunities and realize profits.

\begin{tikzpicture}
\node [mybox] (box){%
    \begin{minipage}{\columnwidth}
    {
    Turning Signals into Dollars: Grok Masters the Art, While DeepSeek Hesitates at Crucial Moments!
    }
    \end{minipage}
};
\node[fancytitle, rounded corners] at (box.south) {{\bf Q3 Takeaway}};
\end{tikzpicture}


\subsection{Trading Personalities Unveiled: Profiling the LLM Investor (Q4)}

\begin{figure}[t!]
    \vspace{-2em}
    \centering
    \includegraphics[width = \textwidth]{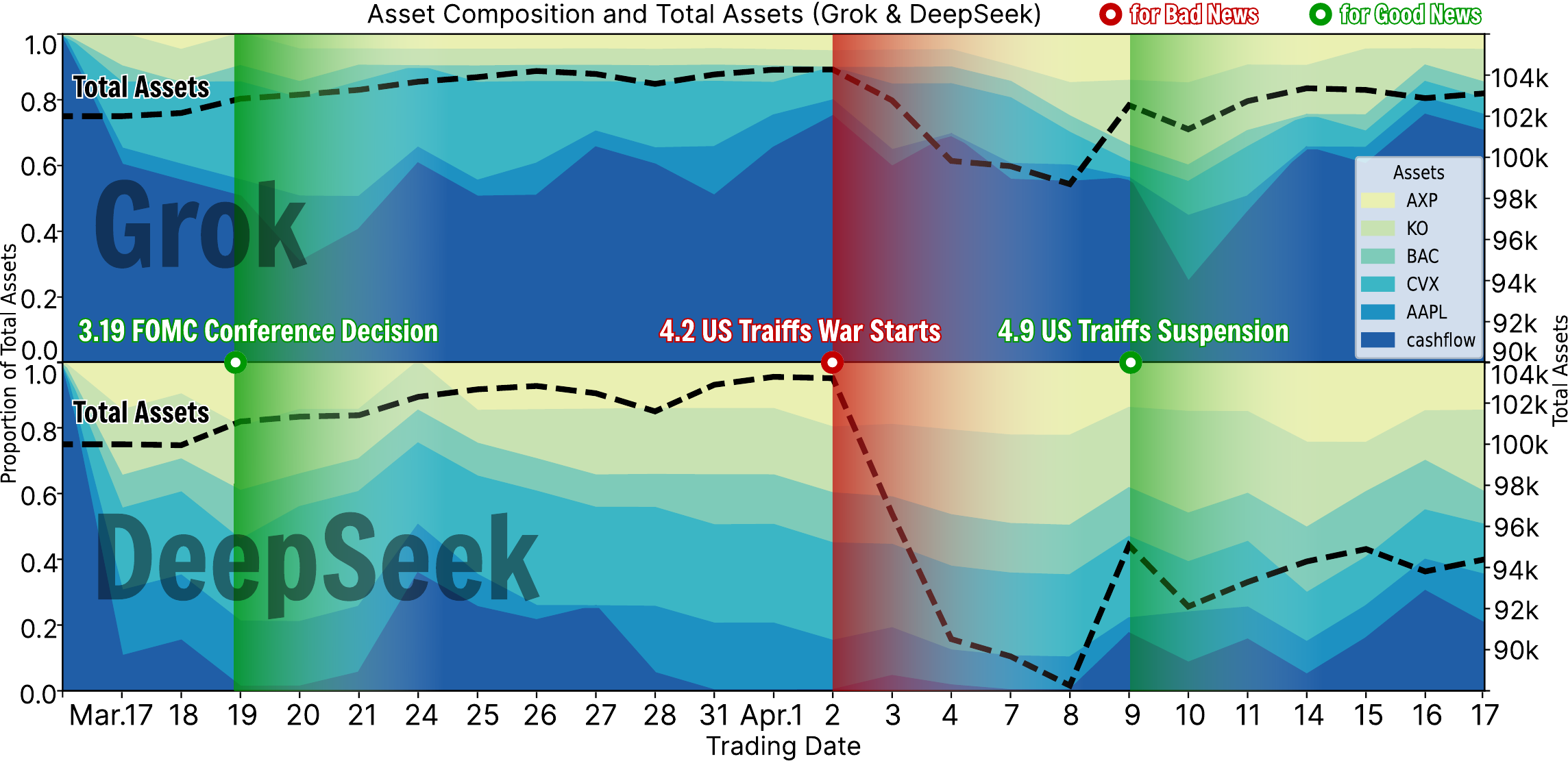}
    \caption{Composition portfolio value for DeepSeek and Grok during the trading period. It shows the holding value of each ticker in the portfolio and the remaining cash flow.}
    \label{fig:composition_N_assets_grok&depsk}
    \vspace{-1em}
\end{figure}

Figure~\ref{fig:composition_N_assets_grok&depsk} illustrates the distinct trading profiles of Grok and DeepSeek through their portfolio compositions and asset evolution during the trading period.
Their trading styles diverged significantly since Day One. Grok initially allocated about 40\% cash to establish positions, maintaining a relatively 60\% high reserve and gradually increasing equity holdings. In stark contrast, DeepSeek aggressively invested nearly 90\% of its initial cash, keeping cash levels consistently below 40\%, indicating high capital utilization.
Subsequently, the dovish remarks made by the FOMC on March 19 brought substantial gains to Deepseek, and this upward trend continued until April 2. However, this low cash reserve severely hampered DeepSeek's flexibility during the market downturn starting April 3, when the US launched a tariff war, impeding timely loss mitigation (refer to case study in Appendix~\ref{appendix:case_2} for details). Grok, with its higher cash position, demonstrated better risk diversification and adaptability. A sufficient cash reserve enables Grok to seize genuine opportunities, significantly increase positions after a sharp decline, and achieve substantial profits during the subsequent rebound after April 9, when US government announced tariffs suspension for most countries.

The models also showed different preferences in sector exposure. Before the period of policy-induced volatility, Grok favored energy (CVX) and consumer staples (KO). DeepSeek, however, concentrated heavily on energy and financial stocks (CVX, BAC, AXP). This lack of sectoral diversification left DeepSeek highly vulnerable to policy shocks without adequate hedging, exacerbating losses during tariff-driven market declines. While Grok mitigated losses by reducing exposure to high-risk assets.

Grok pursued a low-frequency trading strategy with minimal portfolio churn, preferring long-term holdings in what it identified as undervalued blue-chip stocks like KO and CVX. Its maximum drawdown was a mere 3\%, reflecting effective risk management via diversification and dynamic rebalancing. DeepSeek, conversely, adopted a high-frequency, momentum-driven approach, frequently adjusting its portfolio to chase short-term fluctuations. While this initially allowed DeepSeek to profit from selling AAPL at a peak, its portfolio concentration and aggressive cash utilization proved detrimental during the later downturn. Operating with low reserves, DeepSeek was forced into unfavorable \texttt{Buy} or \texttt{Sell} positions, preventing loss recovery. Grok's more measured approach, though slower in early profit growth, ensured greater stability and loss minimization. 

In essence, Grok embodied a prudent, long-term oriented strategy akin to professional fund management, characterized by risk control and diversification. DeepSeek, conversely, exhibited traits of a high-frequency retail speculator—concentrated, momentum-driven, and ultimately vulnerable to market shifts, mirroring the common challenges faced by individual traders.

\begin{tikzpicture}
\node [mybox] (box){%
    \begin{minipage}{\columnwidth}
    {
    Battle of the Trading Styles: Grok’s Steady Precision vs. DeepSeek’s Bold Gambles—Cash is King in a Bearish Market!
    }
    \end{minipage}
};
\node[fancytitle, rounded corners] at (box.south) {{\bf Q4 Takeaway}};
\end{tikzpicture}



\section{Related Work}

\stitle{Benchmarking LLMs in Financial Domain.}
With the blossoming research on LLMs in finance~\cite{xiao2024tradingagents,li2024cryptotrade,li2025hedgeagents}, numerous benchmarks have been developed to evaluate their capabilities in financial contexts. Financial LLM benchmarks have evolved from document-understanding frameworks like TAT-QA~\cite{Zhu2021TATQAAQ}, FinanceBench~\cite{islam2023financebench}, CFBenchmark~\cite{lei2023cfbenchmark}, and FinNLI~\cite{magomere2025finnli} to investment decision-making evaluations such as FinRL-Meta~\cite{liu2022finrl_meta}, FinBen~\cite{xie2024finben} and InvestorBench~\cite{li2024investorbench}. 
These benchmarks share a common focus on assessing LLMs' financial knowledge and reasoning capabilities, but still face a fundamental challenge that the temporal mismatch between model pre-training data and the evaluation window leads to either information leakage when models are tested on historical data they've been trained on, or incomplete evaluation when tested on periods beyond their knowledge boundary. 
Our work shifts from static evaluation to dynamic evaluation to tackle this challenge.

\stitle{Live Benchmarking.}
There has been a growing interest in developing live benchmarks for AI systems to eliminate the ``time-travel'' problem in recent years. Several works have explored contamination-free evaluation approaches in general domains, such as LiveCodeBench~\cite{jain2024livecodebench}, ForecastBench~\cite{karger2024forecastbench}, and LiveBench~\cite{colin2025livebench}. These benchmarks are designed to evaluate LLMs with in-context learning, code generation, and domain-specific tasks, with regular updates to ensure the benchmark is always up-to-date.
Particularly in the financial domain, FinRL-Meta~\cite{liu2022finrl_meta} builds benchmarks for reinforcement learning approaches, and ForecastBench~\cite{karger2024forecastbench} explores the evaluation of LLMs' forecasting capabilities through question-answering that covers market-related questions.
With the advancement of multi-agent systems, several works~\cite{wang2024benchmark,tong2025ide} have developed a self-evolving multi-agent framework for dynamic evaluation on LLMs.
However, in the specific domain of stock trading and fund investment, existing benchmarks have largely remained static, relying on historical data and back-testing approaches. Our work represents the first work that achieves true live benchmarking for fund investment.

\section{Conclusion}

\sys is a novel benchmarking tool for evaluating and comparing the performance of various LLMs in the context of real-time fund investment. It provides a standardized multi-agent trading workflow with a connection to a live environment and LLM factory. Besides, we have conducted empirical studies to show the effectiveness of our framework and reveal the notable potential of LLMs in fund investment. 
Overall, \sys creates a new paradigm for evaluating LLMs in fund investment, which could contribute to the development of reliable and effective financial AI tools.

\section{Limitation}
\label{sec:limitation}
The current implementation highly simplifies the trading context (\ie US stock market only) and does not account for many practical considerations such as transaction fees, market trading restrictions, and hybrid trading strategies. These details could potentially impact trading performance, but have not been evaluated in our current framework. Therefore, we aim to incorporate index-aligned universes, realistic execution frictions, and broader market conditions in future iterations.
Meanwhile, the analysis depth can be further improved by adding more index-based (\eg relative performance to market indices) and LLM-based (\eg reasoning cost, consistency, explainability) evaluation metrics.
Lastly, the evaluation period was short and occurred during a volatile market, which could skew our results toward specific trading approaches. Conducting longer tests across various circumstances (\eg bullish, bearish and volatile markets) would manifest more reliable and applicable insights into LLMs' investment abilities.

\section{Broader Impacts}
\label{sec:impacts}
Our work contributes to the understanding and evaluation of LLMs in financial applications, potentially leading to more robust and effective AI-driven investment strategies for research and educational purposes. However, the application of AI in finance carries inherent risks, including the potential for exacerbating market volatility, introducing biases leading to unfair outcomes, or generating financial losses if misused outside of a controlled research setting. We emphasize that \sys is intended for academic study and benchmarking, and responsible use is paramount to mitigate these risks.

\section{Ethical Statement}
\label{sec:ethics}
The authors take full responsibility for the development of \sys, ensuring that the code repository is publicly available and shared under the MIT license, requiring users to adhere to its terms. \sys is intended for academic and educational purposes only and is not a substitute for professional advice. While efforts have been made to ensure its accuracy, the authors and their institutions disclaim liability for any outcomes arising from its use. Users agree to take responsibility for ethical and lawful use and to indemnify the authors and their affiliates against any claims or damages resulting from reliance on this material.

\section{Acknowledgements}
This paper was supported by
Young Talent Support Project of Guangzhou Association for Science and Technology (QT-2025-001); the NSF of China (62402409); Guangdong Basic and
Applied Basic Research Foundation (2023A1515110545); Guangzhou
Basic and Applied Basic Research Foundation (2025A04J3935);
Guangzhou-HKUST(GZ) Joint Funding Program (2025A03J3714); and
Guangdong Provincial Project (2023CX10X008).
Lastly, we genuinely thank the reviewers for their valuable comments and suggestions during the rebuttal period.


\bibliographystyle{plainnat} 
\bibliography{main}


\newpage
\appendix

\section{Additional Experiments}
\label{appendix:add_exp}

\subsection{Trading Performance by Market Sector}

\begin{figure}[h!]
    \centering
    \subfloat[DeepSeek-V3]{\includegraphics[width=0.49\textwidth]{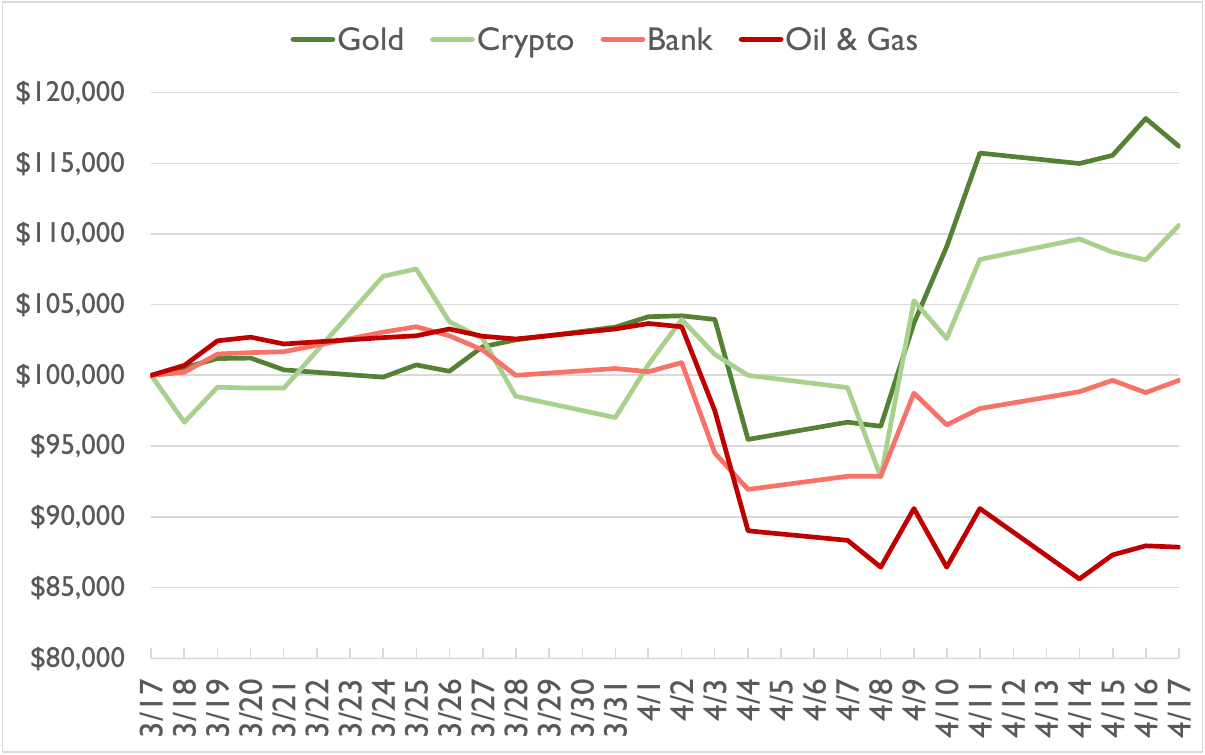}}
    \hfill
    \subfloat[GPT-4.1]{\includegraphics[width=0.49\textwidth]{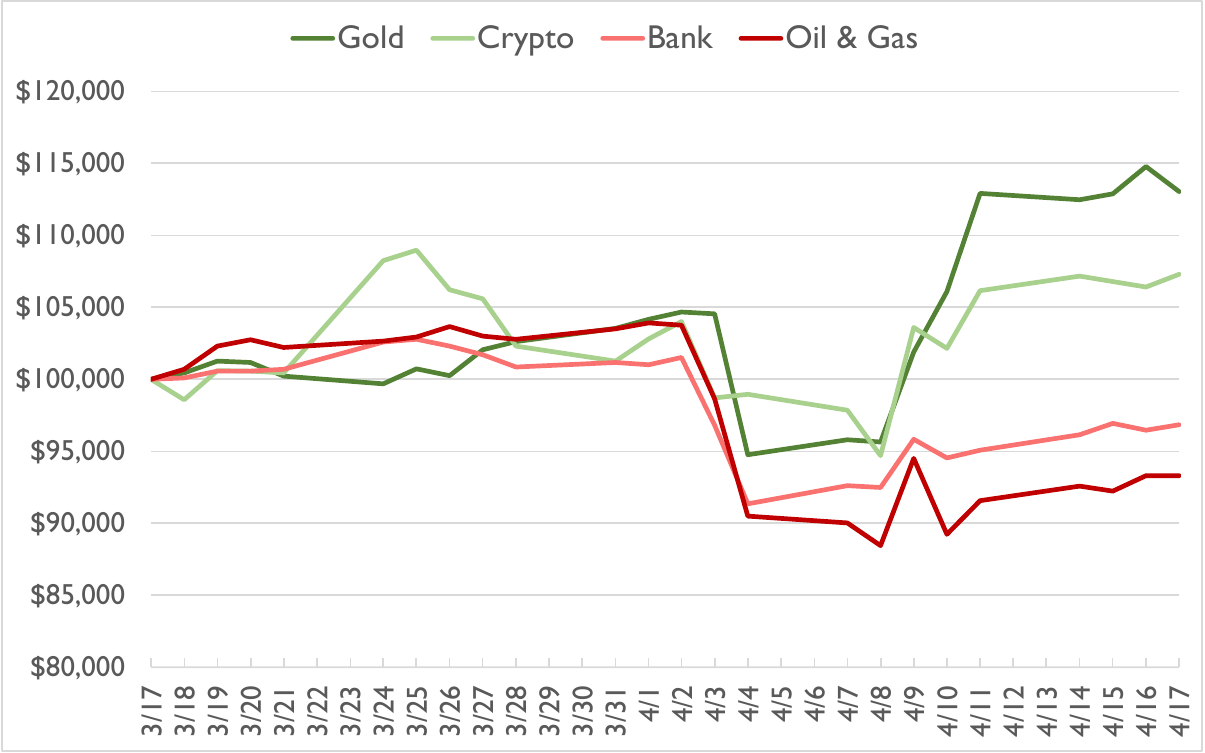}} 
    \caption{Trading performance by market sector. The green and red line indicate a profit gain and loss, respectively.}
    \label{fig:sector_performance}
\end{figure}

In this experiment, we evaluate the investment performance of GPT-4.1 and DeepSeek-V3 in the following four sectors with relevant tickers:
\textbf{Gold} (NEM, GLD, AEM, GFI, KGC), \textbf{Oil \& Gas} (XOM, CVX, NFG, CRGY), \textbf{Crypto} (COIN, MSTR, MARA), \textbf{Banking} (JPM, BAC, WFC, C, RY).

The results are presented in Figure~\ref{fig:sector_performance} with the same trading period in main experiments. Overall, GPT-4.1 exhibits stable growth with low volatility, suitable for conservative strategies. In contrast, DeepSeek-V3 shows high return potential but greater fluctuations, suitable for more aggressive investors. 
Specifically, the Gold and Crypto markets are profitable sectors, while the Oil \& Gas and Banking markets suffer losses. In the profitable sectors, DeepSeek-V3 shows higher returns and a stronger growth potential than GPT-4.1. On the other hand, GPT-4.1 has fewer losses in the Oil \& Gas industry, demonstrating better risk control, while DeepSeek-V3 manifests fewer losses in the Banking sector, indicating better resilience against market downturns.


\subsection{The Cost Efficiency on OpenAI Family}

\begin{figure}[h!]
    \centering
    \includegraphics[width=.6\textwidth]{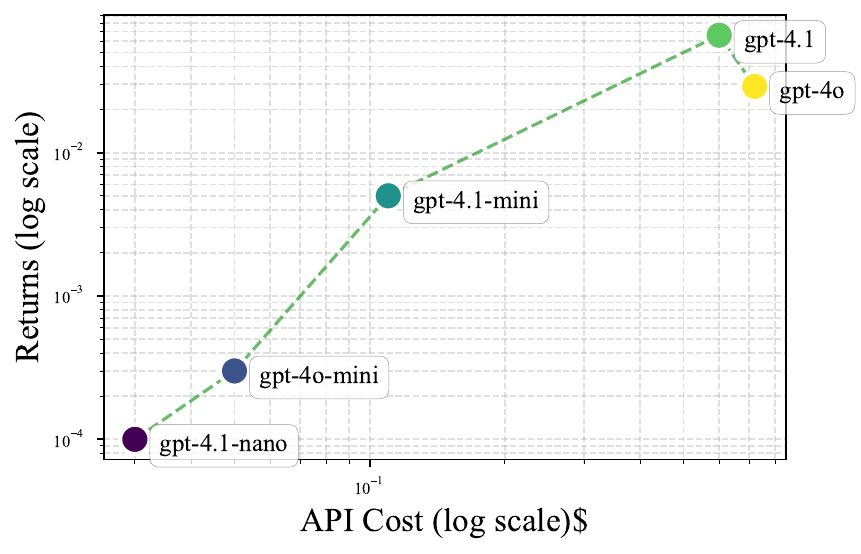}
    \caption{Cost-performance tradeoff analysis for OpenAI family models.}
    \label{fig:gptfamily}
\end{figure}

In the mid April 2025, OpenAI released a new series of GPT-4.1 models~\cite{openai2025gpt41}. In this experiment, we explore the cost-efficiency of the OpenAI family models in our framework. As shown in Figure~\ref{fig:gptfamily}, we evaluate it using two key metrics: LLM API inference cost and trading cumulative return, respectively in x-axis and y-axis. We adopt a shorter trading period from 2025-04-07 to 2025-04-21 on ticker portfolio of AAPL, AXP, BAC, KO, AMT.
We recognize \texttt{gpt-4.1-mini} as the best cost-efficiency (\ie moderate return per dollar spent) choice for most fund applications. For high-frequency trading, we recommend \texttt{gpt-4.1} as it provides the highest return per dollar spent.



\subsection{Extended Trading Performance to entire Q2 2025}

To address the concern on the narrow evaluation window, we expanded our live trading window to cover the entire Q2 2025 (see Table~\ref{tab:q2_perf}). Each LLM continues to actively manage its portfolio and surf in the market, with Grok-3 still maintaining its leading profitability. Some LLMs such as GPT 4.1, Claude 3.7, and DeepSeek V3 achieved net profits eventually, while other LLMs such as GLM 4, Qwen Max, and Gemini 2.5 still incur losses. Results from such a longer period remain consistent with our original findings, which further supports the robustness of our conclusions.

\begin{table}[h]
    \caption{Averaged Weekly Asset Total Value by LLM in Q2 2025. Dates are the end of each week.}
    \label{tab:q2_perf}
    \centering
    \resizebox{\columnwidth}{!}{
    \begin{tabular}{@{}llllllllll@{}}
    \toprule
    Date                  & GPT 4.1   & Claude 3.7 & Grok 3             & Llama 4   & Gemini 2.5 & DeepSeek V3 & Qwen Max & Doubao 1.5 & GLM 4    \\ \midrule
    2025-04-06            & 100207.56 & 99370.47   & \textbf{101031.59} & 100200.65 & 99819.65   & 99305.63    & 99973.54 & 99387.25   & 99629.81 \\
    2025-04-13            & 92824.93  & 93933.56   & \textbf{98978.32}  & 94181.13  & 96280.95   & 91682.35    & 92652.86 & 91204.98   & 91695.78 \\
    \textit{Below are new results} &           &            &                    &           &            &             &          &            &          \\
    2025-04-20            & 94322.82  & 96202.42   & \textbf{101156.52} & 95466.44  & 98212.7    & 94351.44    & 93099.94 & 92098.54   & 92523.98 \\
    2025-04-27            & 94993.84  & 97186.15   & \textbf{101527.47} & 96768.25  & 98205.38   & 96133.87    & 94084.57 & 93030.82   & 93893.19 \\
    2025-05-04            & 96627.35  & 98242.37   & \textbf{102151.96} & 98289.12  & 98349.17   & 97987.88    & 94765.69 & 94182.54   & 95209.34 \\
    2025-05-11            & 96247.15  & 98319.67   & \textbf{102227.45} & 97770.66  & 97374.24   & 98223.28    & 94376.73 & 93827.1    & 95528.87 \\
    2025-05-18            & 98815.25  & 100726.09  & \textbf{103425.42} & 102377.7  & 98323.0    & 101482.47   & 95853.75 & 97527.84   & 98038.4  \\
    2025-05-25            & 97701.93  & 99340.52   & \textbf{103385.97} & 100680.71 & 97822.53   & 99767.74    & 94142.15 & 96939.64   & 96392.79 \\
    2025-06-01            & 98589.52  & 99419.7    & \textbf{103625.78} & 100703.5  & 98473.83   & 99969.97    & 94540.49 & 97865.15   & 96501.95 \\
    2025-06-08            & 99292.85  & 99914.3    & \textbf{103972.52} & 101252.86 & 98746.37   & 100703.84   & 94868.62 & 98584.24   & 96893.33 \\
    2025-06-15            & 99942.06  & 100360.74  & \textbf{104263.5}  & 101935.87 & 99192.8    & 101416.86   & 95380.01 & 99262.71   & 97060.48 \\
    2025-06-22            & 99470.03  & 99668.47   & \textbf{104700.82} & 101698.64 & 98919.57   & 100873.6    & 95216.37 & 98985.76   & 96356.1  \\
    2025-06-29            & 101442.76 & 101234.08  & \textbf{106262.44} & 102859.48 & 99762.68   & 102851.0    & 96383.33 & 100865.84  & 97670.72 \\ \bottomrule
    \end{tabular}
    }
    \end{table}
\newpage
\section{Technical Details}

\subsection{Data Models and Operational Schemas}
\label{appendix:schema}
To facilitate the message communication and information flow throughout the system, we implement hierarchical data models and operational schemas that standardize agent interactions. These elements ensure consistent information processing while maintaining semantic richness across the multi-agent framework. 

Our core data models encapsulate domain-specific financial information, not limited to the following examples:
\textbf{MediaNews} are normalized containers for company-specific news, press releases, and policy updates; \textbf{Insiders} are formalized tracking of insider transactions, executive changes, and corporate governance events; \textbf{Fundamentals} are standardized models of financial statements, valuation metrics, and growth indicators; \textbf{OHLCV} metrics represent the daily trading statistics, contributing to the calculation of technical indicators. All data models are designed to be the information upstream of the specialised agents.

Apart from data models, we also define a set of operational schemas that govern system behavior: \textbf{Signal}: Structured output from analyst agents containing direction (Bullish, Bearish, Neutral) and detailed justification; \textbf{Decision}: Formalized investment actions (Buy, Sell, Hold), number of shares, price that day, and reasonings; \textbf{Portfolio}: Comprehensive state representation tracking cashflow, and holding positions with corresponding shares and risk exposure. Notably, all of the above schemas are encapsulated into a unified object, \textbf{FundState}, which is a system-wise message container that persists the current state of the fund.

If the Policy and Fundamental analysts are selected, they will receive upstream \textbf{MediaNews} and \textbf{Fundamentals} as input, and output corresponding \textbf{Signal} object. Consequently, the portfolio manager will analyse based on those signals and current holding positions, and output a \textbf{Decision} object. Finally, the \textbf{Portfolio} object will be updated according to the \textbf{Decision} object. Eventually, this model-schema governance enables both specialized analysis and integrated decision-making while maintaining strict data consistency and provenance tracking throughout the system.

\subsection{Memory Management}
\label{appendix:memory}

Memory management is crucial for maintaining context and learning in our multi-agent system~\cite{wang2024agent,liu2025advances}. We implement a dual-memory architecture that combines short-term operational memory with long-term historical memory to enable both immediate decision-making and continuous learning.

\stitle{Short-term Memory.}
The primary short-term memory in our system is implemented through the \textbf{FundState} object, which serves as a thread-scoped memory container~\cite{kim2023machine,peng2023check}. This stateful object maintains the current operational context of the fund, encompassing current portfolio positions and cash balance, recent trading decisions and their rationales, active signals from analyst agents, and the latest market data and news context. The \textbf{FundState} is updated in real-time as the system processes new information and makes decisions, ensuring all agents have access to the most recent operational context. This short-term memory is essential for maintaining consistency across agent interactions and enabling coherent decision-making within a single trading session.

\stitle{Long-term Memory.}
Our system maintains long-term memory through comprehensive trading history records from live market interactions~\cite{cowan2008differences,atkinson1968human}. This historical memory serves as a foundation for performance tracking and analysis of trading strategies, enabling the system to learn from past decisions and their outcomes. Through pattern recognition across different market conditions, the system continuously improves its decision-making capabilities. This persistent memory layer helps the system adapt and improve its performance over time by incorporating lessons from previous trading sessions.

The combination of short-term operational memory through \textbf{FundState} and long-term historical memory creates a robust memory architecture that supports both immediate decision-making and continuous system improvement. This dual-memory approach enables our multi-agent system to maintain context awareness while learning from past experiences.

\subsection{Evaluation Interface}
\label{appendix:interface}
The interface serves as the first AI Live Investment Arena, designed to evaluate the trading and investment capabilities of various Large Language Models (LLMs) across financial markets. It provides a comprehensive environment for assessing how LLMs can ingest financial information, drive multi-agent systems, and make trading decisions in real-world market scenarios. You can explore its features further by visiting \url{https://deepfund.paradoox.ai/}. The main features are:

\begin{table}[h!]
    \centering
    \label{tab:interface_features}
    \caption{Key features of the DeepFund valuation interface.}
    \begin{tabular}{p{0.2\linewidth}p{0.75\linewidth}}
    \toprule
    \textbf{Feature Name}          & \textbf{Description}                                                                                                            \\ \midrule
    Performance Leaderboard        & A competitive ranking system comparing LLM models based on investment metrics (total returns, daily returns, portfolio values). \\
    Interactive Data  Visualization & Charts displaying cumulative returns over time with adjustable periods for detailed performance analysis.                       \\
    Portfolio Analysis             & Detailed breakdown of each LLM agent's portfolio, including holdings, asset allocation, and value distribution.                 \\
    Market Comparison              & Direct comparison between LLM performance and major market indices (NASDAQ, S\&P 500, DOW JONES).                               \\
    Agent Lab                      & An environment for users to customize and develop their own LLM trading agents to compete.                                      \\
    Reports Section                & Provides analytical reports on performance and market trends.                                                                   \\ \bottomrule
    \end{tabular}
    \end{table}

\subsection{Parameter Settings}
\label{appendix:parameter}

\begin{table}[h!]
    \centering
    \label{tab:parameter_settings}
    \caption{\sys parameter settings.}
    \begin{tabular}{@{}ccc@{}}
    \toprule
    \textbf{Parameter}       & \textbf{Default Value} & \textbf{Usage} \\ \midrule
    LLM temperature          & 0.5                    & Control the randomness of the LLM inference. \\
    Retry times              & 3                      & Number of retries for LLM inference. \\
    Technical window        & 100                     & OHLCV covered days for technical analysis. \\
    Insider count          & 10                      & Insider transactions for insider analysis. \\
    Number of news           & 10                     & Compnany news and policy analysis. \\
    Decision memory size         & 5                    & Number of past recent actions for decision-making. \\ \bottomrule
    \end{tabular}
\end{table}

\subsection{Systematic Scalability}
\label{appendix:scalability}
To ensure long-term viability and extensibility, DeepFund implements a modular architecture that decouples core functionalities into components. This design enables seamless integration of diverse LLMs, data sources, and agent composition without architectural modifications. \textbf{LLM module} provides a unified interface that abstracts provider-specific implementations, enabling fair comparison and rapid integration of new LLMs. \textbf{Data source module} implements a similar abstraction for financial information, standardizing diverse sources into consistent internal data models. \textbf{Agent composition module} enables community extension through a well-defined protocol for adding specialized analytical methodologies. Thus, researchers can contribute novel analytical approaches that seamlessly integrate with the existing system. This interface architecture transforms DeepFund from a fixed evaluation platform into an evolving research ecosystem, providing standardized benchmarking while supporting continuous incorporation of advances in LLM technology, financial data analysis, and agent system design.

\newpage
\section{Signal and Decision Statistics}
\label{appendix:key_stats}

\begin{figure}[h!]
    \centering
    \includegraphics[width=.75\textwidth]{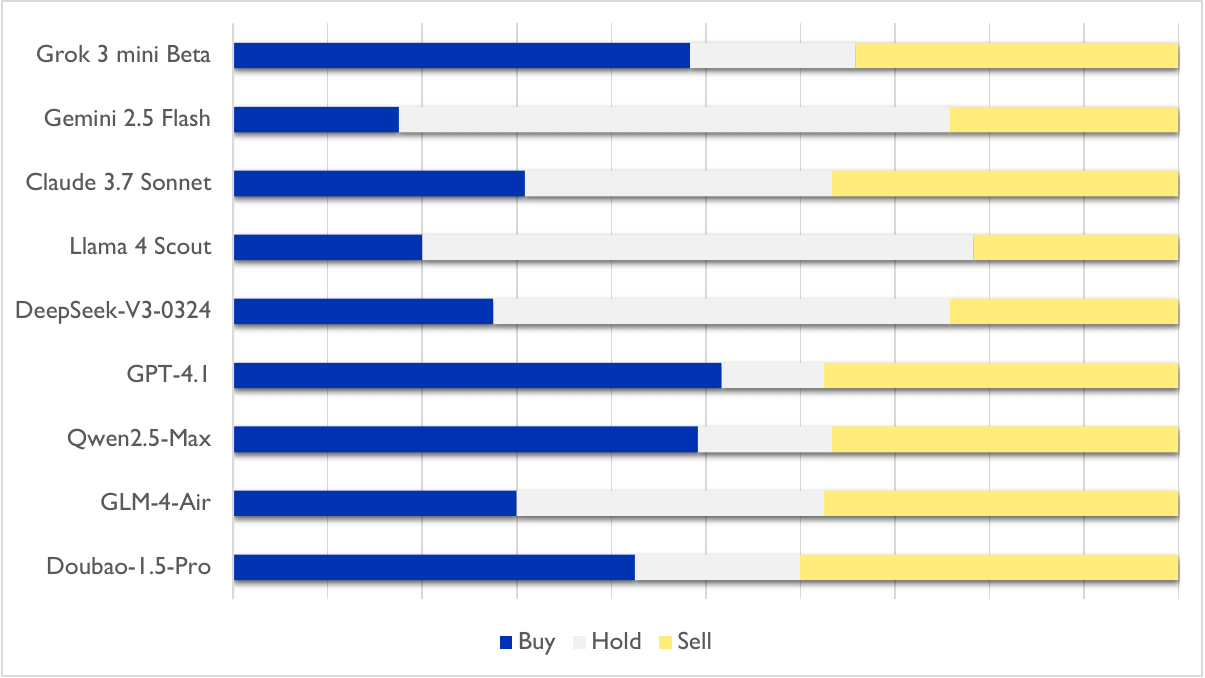}
    \caption{Decision statistics by LLMs across all tickers.}
    \label{fig:decision_stats}
\end{figure}

\begin{figure}[h!]
    \centering
    \subfloat{\includegraphics[width=0.48\textwidth]{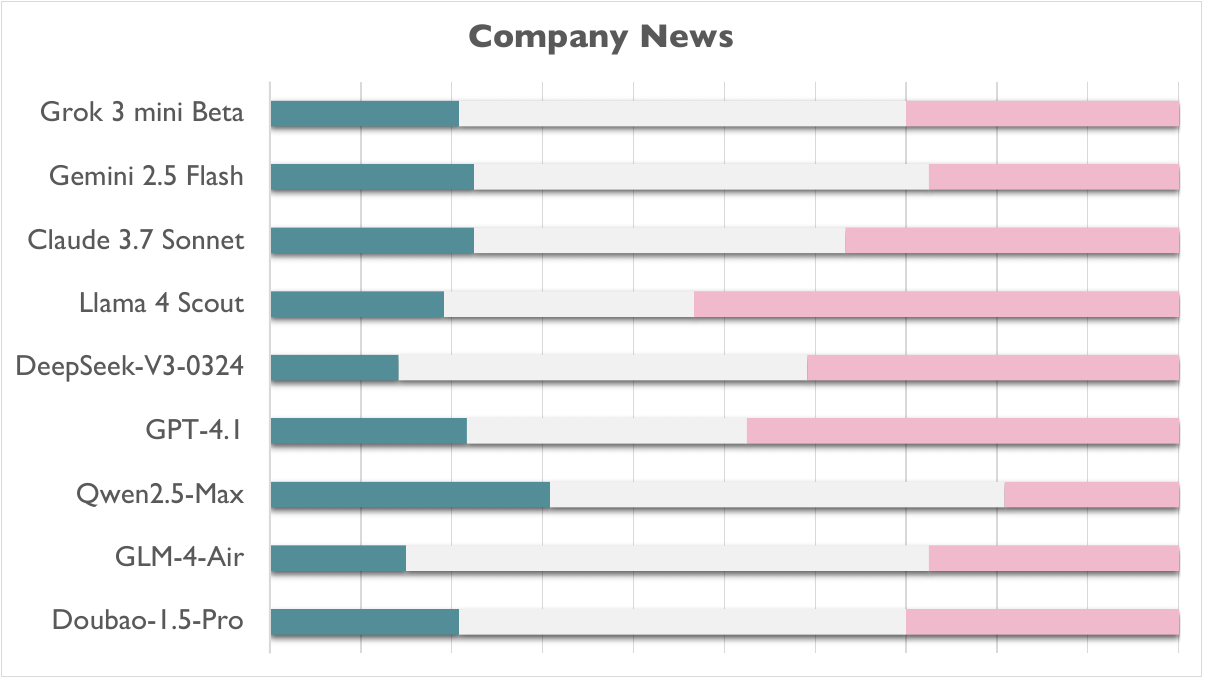}}
    \hfill
    \subfloat{\includegraphics[width=0.48\textwidth]{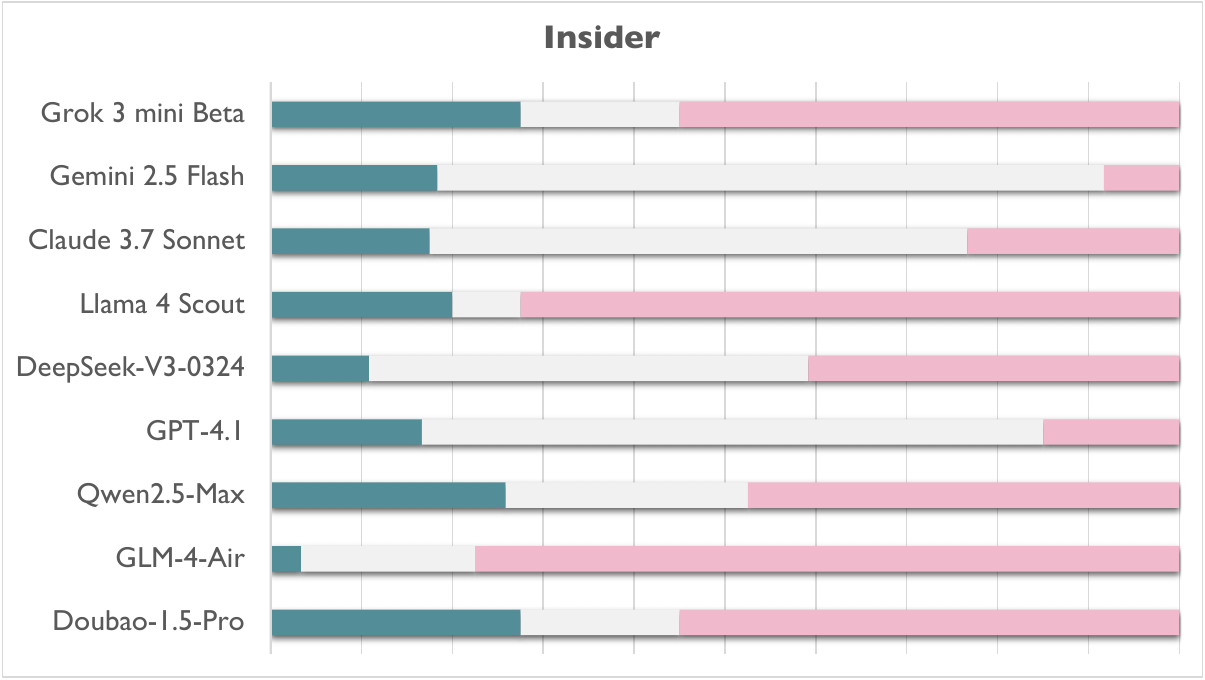}} \\
    \subfloat{\includegraphics[width=0.48\textwidth]{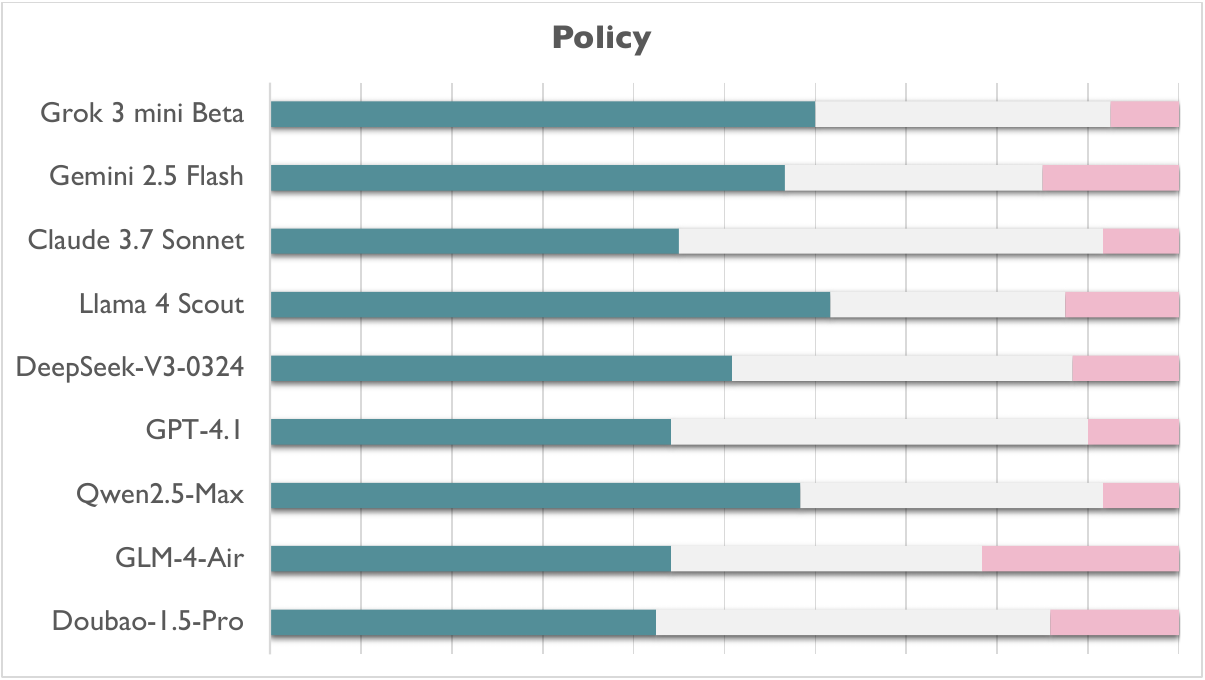}}
    \hfill
    \subfloat{\includegraphics[width=0.48\textwidth]{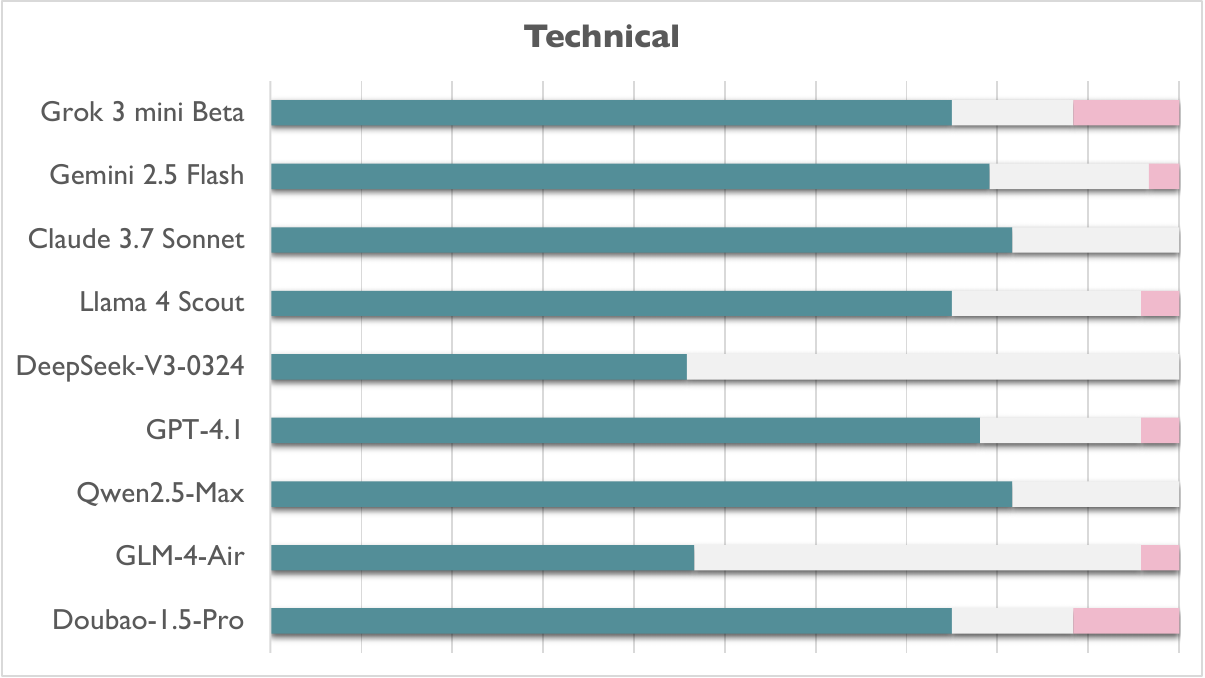}}
    \caption{Analytical signal statistics by LLMs across all tickers.}
    \label{fig:signal_stats}
\end{figure}

As shown in Figures~\ref{fig:decision_stats} and~\ref{fig:signal_stats}, the various LLMs exhibit distinct decision-making behaviors and analytical signal processing patterns. The correlation between overt trading behaviors (i.e., buy, sell, and hold decisions) and the corresponding signal data offers valuable dimensions for a more nuanced analysis and characterization of the distinct trading styles inherent to each model.

\stitle{Decision Distribution.}
We observe that distinct models employ varied signal processing mechanisms and decision-making frameworks, resulting in a spectrum of observable trading strategies. For instance, models characterized by a high proportion of hold recommendations exhibit discernible differences in their underlying signal information profiles when contrasted with models demonstrating more frequent buy or sell activities. 

\stitle{Signal Distribution.}
These models display heterogeneous sensitivities to diverse market signals, with certain signal categories evidently assuming a predominant role in the strategic outputs of particular models. This variance suggests the implementation of sophisticated and potentially model-specific internal weighting systems for signal aggregation and interpretation.

\newpage
\section{Evaluation Metrics}
\label{appendix:metric}

In this section, we provide detailed definitions and formulas for the evaluation metrics used in our experiments.

\stitle{Cumulative Return (CR).}~\cite{hull2012risk}
Total percentage gain or loss from the initial investment:
\begin{equation}
\text{CR}(\%) = \left(\frac{P_{\text{final}}}{P_{\text{initial}}} - 1\right) \times 100%
\end{equation}
where $P_{\text{initial}}$ and $P_{\text{final}}$ denote the initial and final portfolio values, respectively.

\stitle{Cumulative Return at Buy \& Hold ($CR_{bnh}$).}~\cite{fama1988permanent}
Buy \& Hold is a passive investment approach, where an investor purchases stocks and holds onto them for an extended period regardless of market fluctuations. We harness this strategy as an alternative investigation to evaluate the performance of LLMs. The portfolio is initialized by LLM since day 1 and held until the end of the test period.

\stitle{Sharpe Ratio (SR)}~\cite{sharpe1994sharpe}
Excess return divided by its standard deviation, using the risk-free rate based on the 1-month US Treasury bill (4.29\% as of 2025-04-17):

\begin{equation}
\text{SR} = \frac{\overline{r_e}}{\sigma_{r_e}} \times \sqrt{252}
\end{equation}

where $\overline{r_e}$ is the average daily excess return ($r_s - r_f$), $\sigma_{r_e}$ is the standard deviation of excess returns, and 252 is the number of trading days in a year.

\stitle{Maximum Drawdown (MDD)}~\cite{magdon2004maximum}
Largest percentage decline from peak portfolio value, indicating downside risk:
\begin{equation}
\text{MDD}(\%) = \max_{t \in [0,T]} \left(\max_{s \in [0,t]} \frac{P_s - P_t}{P_s}\right) \times 100%
\end{equation}
where $P_t$ is the portfolio value at time $t$, and $T$ is the total investment horizon.

\stitle{Win Rate (WR)}~\cite{chan2009quantitative}
Percentage of profitable trades executed:
\begin{equation}
\text{WR}(\%) = \frac{\sum_{i=1}^{n} \mathbf{1}{r_i > 0}}{n} \times 100%
\end{equation}
where $r_i$ is the return of the $i$-th trade, $\mathbf{1}{r_i > 0}$ is the indicator function that returns 1 if the trade was profitable, and $n$ is the total number of trades.

\stitle{Beta ($\beta$)}~\cite{jensen1968performance}
Portfolio volatility relative to the S\&P 500:
\begin{equation}
\beta = \frac{\text{Cov}(r_s, r_m)}{\text{Var}(r_m)}
\end{equation}
where $r_s$ is the return of the strategy, $r_m$ is the return of the market (S\&P 500), $\text{Cov}(\cdot)$ denotes covariance, and $\text{Var}(\cdot)$ denotes variance.

\stitle{Alpha ($\alpha$)}~\cite{jensen1968performance}
Excess return compared to the market benchmark (S\&P 500):
\begin{equation}
\alpha = r_s - [r_f + \beta(r_m - r_f)]
\end{equation}
where $r_s$ is the strategy return, $r_f$ is the risk-free rate(4.29\% as of 2025-04-17), $r_m$ is the market return, and $\beta$ is the strategy's beta as defined above.
\newpage
\section{Prompt Template}
\label{appendix:prompt}

As we committed to open source our code repository, the full details are documented in \texttt{src/llm/prompt.py} under our project repository \url{https://github.com/HKUSTDial/DeepFund}. Here we provide the prompt template of technical analyst and portfolio manager for illustration.

\subsection{Technical Analyst}

\begin{tcolorbox}[left=1mm,right=1mm,top=1mm,bottom=1mm,colback=white]
You are a technical analyst evaluating ticker using multiple technical analysis strategies. The following signals have been generated from our analysis:

\medskip
Price Trend Analysis: \textcolor{teal}{\{trend\}} \\
Mean Reversion: \textcolor{teal}{\{mean\_reversion\}} \\
RSI: \textcolor{teal}{\{rsi\}} \\
Volatility: \textcolor{teal}{\{volatility\}} \\
Volume Analysis: \textcolor{teal}{\{volume\}} \\
Support and Resistance Levels: \textcolor{teal}{\{price\_levels\}}

\medskip
You must provide your analysis as a structured output with the following fields:
\begin{itemize}[leftmargin=*]
    \item signal: One of \textcolor{violet}{[``Bullish'', ``Bearish'', ``Neutral'']}
    \item justification: A brief explanation of your analysis
\end{itemize}

Your response should be well-reasoned and consider all aspects of the analysis.
\end{tcolorbox}

Specifically, the technical analyst process the OHLCV data to analyse price patterns, momentum indicators, and statistical trends.

\subsection{Portfolio Manager}

\begin{tcolorbox}[left=1mm,right=1mm,top=1mm,bottom=1mm,colback=white]
You are a portfolio manager making final trading decisions based on decision memory, and the provided optimal position ratio.

\medskip
Here is the decision memory:
\textcolor{teal}{\{decision memory\}}

\medskip
Current Price: \textcolor{teal}{\{current price\}} \\
Holding Shares: \textcolor{teal}{\{current shares\}} \\
Tradable Shares: \textcolor{teal}{\{tradable shares\}}

\medskip
If the value of tradable shares is positive, you can buy more shares.\\
If the value of tradable shares is negative, you can sell some shares.\\
If the value of tradable shares is close to 0, you can hold.

\medskip
You must provide your decision as a structured output with the following fields:
\begin{itemize}[leftmargin=*]
\item action: One of \textcolor{violet}{[``Buy'', ``Sell'', ``Hold'']}
\item shares: Number of shares to buy or sell, set 0 for hold
\item price: The current price of the ticker 
\item justification: A brief explanation of your decision
\end{itemize}

Your response should be well-reasoned and consider all aspects of the analysis.
\end{tcolorbox}

Specifically, the decision memory is retrieved from trading history. The current price represents the latest price of the ticker, holding shares represent the shares currently held based on the current fund state. The tradable shares represent the shares that can be bought or sold, which is derived from the risk control module.

\newpage
\section{Case Studies}

\subsection{Divergent Signals from Policy News on April 10th}
\label{appendix:case_1}


\begin{tcolorbox}[left=1mm,right=1mm,top=1mm,bottom=1mm,colback=white]
\textbf{Fiscal View:}
\begin{itemize}[leftmargin=*]
\item Beijing white paper outlines China trade stance before retaliatory strike on US.
\item  `Brace for impact': Chinese economist warns the gloves are off in US trade war.
\item United Airlines, Microchip Technology lead the stocks posting 20\%+ reversals on tariff pause.
\end{itemize}

\medskip
\textbf{Monetary View:}
\begin{itemize}[leftmargin=*]
\item Bond Market Twists Tariff Arm to 90 Day Pause.
\item Citigroup Vs Wells Fargo: Which Bank Stock is a Smarter Investment?
\item Here's How to Play Goldman Stock Ahead of Its Q1 Earnings Release.
\end{itemize}

\rule{\textwidth}{1pt}

\medskip
\textbf{Grok:} \\
\textbf{Signal:} \textcolor{magenta!60}{Bullish} \\
\textbf{Justification:} The news highlights a tariff pause leading to stock surges,increased investor confidence, and positive market reactions in stocks and crypto, outweighing short-term uncertainties from trade tensions, suggesting potential economic growth over the next 6 months. 

\medskip
\textbf{DeepSeek:} \\
\textbf{Signal:} \textcolor{teal}{Bearish} \\
\textbf{Justification:} The fiscal policy news highlights escalating trade tensions between China and the US, including retaliatory measures and warnings from Chinese economists. This uncertainty could negatively impact global trade and economic stability in the short term. Additionally, while some positive developments like tariff pauses and bank stock surges are noted, the overall tone suggests heightened risks and volatility.

\end{tcolorbox}

\subsection{Decision Discrepancy on April 3rd}
\label{appendix:case_2}

\begin{tcolorbox}[left=1mm,right=1mm,top=1mm,bottom=1mm,colback=white]

\textbf{Trading Facts:}
\begin{itemize}[leftmargin=*]
\item Ticker: Chevron Corp (NYSE: CVX)
\item Price of the Day: 156.12
\item Historical Prices In The Past Four Days: [166.47, 168.51, 167.29, 166.06]
\end{itemize}

\rule{\textwidth}{1pt}

\medskip
\textbf{Grok:} \\
Decisions In The Past Four Days: [Hold, Hold, Hold, Hold] \\
Holding Shares: 184, Cashflow: 4252.06, Total Asset: 96622.39 \\

\textbf{Action:} \colorbox{yellow!80}{Sell 27 shares} \\
\textbf{Justification:} The tradable shares value is -27.0, indicating an opportunity to sell shares to align the portfolio with the optimal position ratio. The current price of 156.12 is lower than recent historical prices, but selling the excess shares will help rebalance the portfolio.

\medskip
\textbf{DeepSeek:} \\
Decisions In The Past Four Days: [Buy 1 share, Sell 61 shares, Sell 1 share, Buy 30 shares] \\
Holding Shares: 184, Cashflow: 59941.02, Total Asset: 100768.53 \\

\textbf{Action:} \colorbox{blue!50}{Buy 68 shares} \\
\textbf{Justification:} Tradable shares are positive at 68.0, indicating an opportunity to buy more shares based on the optimal position ratio.

\end{tcolorbox}


\end{document}